\documentclass[twocolumn,prb,showpacs,amsmath,amssymb,english]{revtex4-1}
\usepackage{graphicx}
\usepackage{dcolumn}
\usepackage{bm}
\begin{document}
\title{Shape resonance for the anisotropic superconducting gaps near a Lifshitz transition: the effect of electron hopping between layers}
\author{Davide Innocenti}
\author{Sergio Caprara}
\author{Nicola Poccia}
\author{Alessandro Ricci}
\author{Antonio Valletta$^+$}
\author{Antonio Bianconi}
\affiliation{Physics Department, Sapienza University of Rome,
Piazzale Aldo Moro 2, 00185 Rome, Italy}
\date{\today}
\affiliation{$^+$Institute for Microelectronics and Microsystems, IMM CNR, Via del Fosso del Cavaliere 100, 00133 Roma, Italy}
\date{\today}
\begin{abstract}
The multigap superconductivity modulated by quantum confinement effects in a superlattice of quantum wells
is presented. Our theoretical BCS approach captures the low-energy physics of a shape resonance in the
superconducting gaps when the chemical potential is tuned near a Lifshitz transition. We focus on the case
of weak Cooper-pairing coupling channels and strong pair exchange interaction driven by repulsive Coulomb
interaction that allows to use the BCS theory in the weak-coupling regime neglecting retardation effects
like in quantum condensates of ultracold gases. The calculated matrix element effects in the pairing
interaction are shown to yield a complex physics near the particular quantum critical points due to
Lifshitz transitions in multigap superconductivity. Strong deviations of the ratio $2\Delta/T_c$ from
the standard BCS value as a function of the position of the chemical potential relative to the Lifshitz
transition point measured by the Lifshitz parameter are found. The response of the condensate phase to
the tuning of the Lifshitz parameter is compared with the response of ultracold gases in the
BCS-BEC crossover tuned by an external magnetic field. The results provide the description of the
condensates in this regime where matrix element effects play a key role.
\end{abstract}

\pacs{74.62.-c,74.70.Ad,74.78.Fk}

\maketitle

\section{Introduction}

For a long time the conventional theoretical models for high-temperature superconductivity (HTS) in
cuprates have been based on a single-component electronic system with a single effective band, while
compelling experimental evidence of two bands, resulting from different orbitals, crossing the Fermi
level was reported since 1988.\cite{a} Unconventional theories for HTS have been proposed focusing on
the characteristics of multigap superconductivity.\cite{22,22a,23}  The possible role of a Lifshitz
transition\cite{54,55} [called also electronic topological transition (ETT) or quantum phase transitions
(QPT) of the 2.5-order] in HTS was discussed in the frame of the single band model.\cite{55a} In 1993
it was proposed a novel paradigm that identifies the quantum mechanism for rising the critical temperature
in the high-temperature range: the shape resonance in superconducting gaps in multiband systems where
the chemical potential is tuned near a Lifshitz transition in one of the bands.\cite{28,29,29a,15,15a,15b}

In the last ten years the experimental research has been the driving force for shifting the majority of
the scientific community toward the novel paradigm that HTS emerges in complex systems with multiple
electronic components in the proximity of a Lifshitz transition. In fact two electronic components at the
Fermi level have been found by many groups in cuprates,\cite{1,2,3,4,5,6} showing two (or more) strongly
correlated bands with tendency to phase separation.\cite{7,7x} The two bands in YBaCuO system observed in
1988\cite{a} are now confirmed by an increasing number of experiments\cite{77a,77b,77c,77d} and recent
theoretical models consider the new paradigm of a multiband system near a Lifshitz transition.\cite{78,78a}
The theory of multiband or multigap superconductivity in the clean limit was developed since long ago, but
it was accepted only after the discovery of the two-gap superconductivity at high temperature in magnesium
diboride MgB$_2$\cite{8} and in doped diborides.\cite{10,11,12,12a} Later on, low-temperature multiband
superconductivity has been found in several materials, such as titanates, ruthenates, boro-carbides,
selenides. However, the origin of HTS in MgB$_2$ has been assigned by many authors to the strong
electron-phonon intraband coupling in the $\sigma$ band. On the contrary the unique features of multigap
superconductivity near a Lifshitz transition\cite{36,10,12a} received little attention by the majority
of the HTS community. High-temperature multiband  superconductivity has been discovered in 2008 in
Fe pnictide multilayers. These materials show a clear weak intraband electron-phonon coupling, therefore
multigap superconductivity at 50K in pnictides has determined in these last two years the gradual shift
of the scientific opinion toward the nowadays widely accepted idea that specific features of multigap
superconductivity may be essential for the emergence of HTS.\cite{16,16a,17} Recently, the search for HTS
is focusing on the control of multiband superconductivity in carbon nanotubes,\cite{18} graphene
bilayers,\cite{19} and graphane.\cite{20}

Looking for common features of different high-$T_c$ superconducting materials (e.g., cuprates, diborides
and pnictides), one could list the following:
\emph{first}, multigap superconductivity;\cite{24}
\emph{second}, multiple electronic components;\cite{25}
\emph{third}, a heterostructure at the atomic limit, made of superconducting atomic layers intercalated
by spacers made of a different material, forming a superlattice\cite{28,29} as shown in Fig. 1;
\emph{fourth}, the tuning of the Fermi energy (using any physical, chemical or material manipulation
method, such as doping and superlattice misfit strain) in order to reach a particular point in the
electronic band structure.\cite{60,26,27,27a}
These are the features which were proposed to be essential for the synthesis of new high-$T_c$
superconductors, with $T_c$ eventually reaching room temperature, in Refs. \onlinecite{28,29,29a}. The
particular condition yielding high $T_c$ was thereby identified as the tuning of the Fermi energy at
a \emph{shape resonance} for the superconducting gaps near a Lifshitz transition.
In this work we first discuss the fundamental points of shape resonances in multiband superconductors made
of a first two-dimensional (2D) cylindrical Fermi Surface and a second Fermi surface, tuning the
chemical potention near the
Lifshitz transition. We present the results for the a system of weak intraband coupling so the system is
well described by the BCS weak-coupling limit and we consider a case where the interband pairing is the
dominant interaction so that it could describe the case of pnictides where the interband pairing is the
dominant interaction. In this regime we investigate the effect of variable electron hopping between the
layers that is the characteristic feature of superlattices of superconducting layers at atomic limit.

\section{Shape resonances for superconducting gaps}

The concept of \emph{shape resonance} for superconducting gap parameters was first
introduced by Blatt and Thompson for a single 2D membrane\cite{30} and it has been developed for a
single nanowire.\cite{32} The \emph{shape resonance} for the superconducting gaps in superlattices of
quantum wells, (quantum layers, quantum wires and quantum dots)\cite{15,28,29} has been proposed to
yield three-dimensional (3D) superconductors.
Shape resonances in a superlattice of stripes in cuprates,\cite{33,34,35} and pnictides\cite{17}, of
layers in diborides,\cite{36,37,38,39} and in a superlattice of nanowires in carbon nanotubes, have been
discussed.\cite{18} The novel scenario for a multiband system with a dominant role of interband pairing
and negligible intraband coupling has been proposed before the discovery of pnictides.

\begin{figure}[tbp]
\centering
\includegraphics [angle=0,scale=0.52]{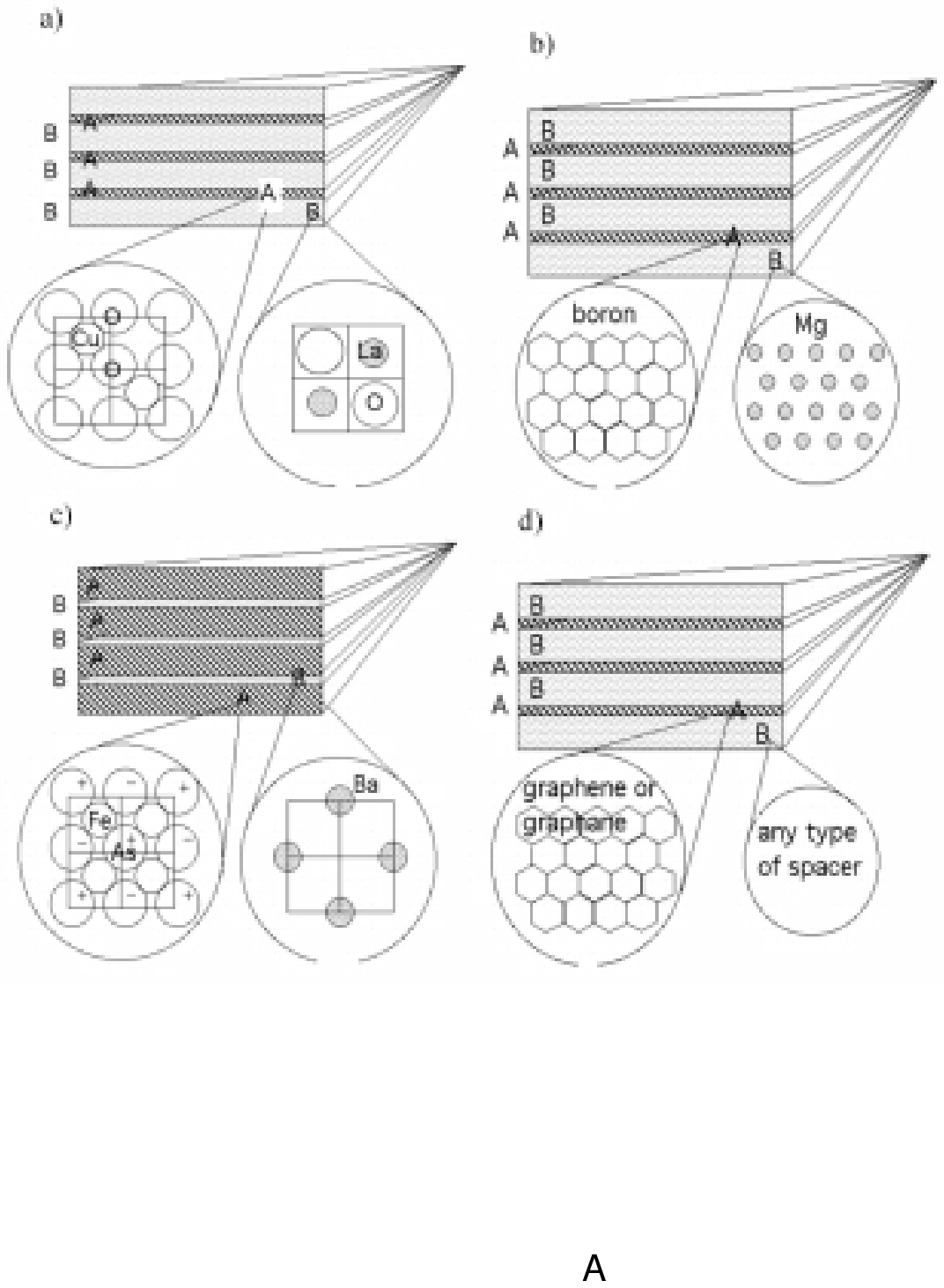}
\caption{(Color online). Pictorial view of artificial heterostructures realized as
superlattices of stacked planes, with particulars of the superconducting layer and
of the spacer layers, in four different families of compounds: (a) cuprates, (b)
magnesium diborides, (c) iron pnictides, and (d) graphene or graphane layers.}
\label{fig1}
\end{figure}

The definition of \emph{shape resonance} in the theory of superconductivity\cite{30} is borrowed
from nuclear physics. Ettore Majorana first in 1929-1930 developed a theoretical model to describe the
artificial disintegration of nuclei by bombardment with $\alpha$-particles.\cite{41}  By following the
quantum dynamics of a state resulting from the superposition of a discrete state with a continuum one,
whose interaction is described by a given potential term, Majorana has made the first application of
the concept of the idea of quasi-stationary states interacting with the continuum.\cite{41,41b} The
theory of shape-elastic scattering cross section for nucleon-nucleus resonances, where the total energy
of incoming or outgoing particles and the quasi stationary states of the compound nucleus are nearly
degenerate, was developed in the fifties by Blatt, Feshbach and de Shalit.\cite{42,42a,43} The
\emph{shape resonance} in nuclear scattering from a potential exhibits characteristic peaks, as a
function of energy, for values of energy such that an integer number of wavelengths fits within the
nuclear potential well. The \emph{shape resonances} have been measured in electron-atom
scattering\cite{44} and electron-molecule scattering.\cite{45,46} The name \emph{Feshbach resonance}
was coined to indicate a particular \emph{shape resonance}, where the quasi-bound state is near zero
energy.\cite{46} The \emph{Feshbach resonance} in atomic association and dissociation processes occurs
in ultra-dilute and ultra-cold atomic gases where the energy of a diatomic molecule (tuned by an
external magnetic field) is degenerate with the chemical potential of the atomic gas.\cite{47} In this
condition the exchange interaction give the Feshbach resonance. The discovery that in boson and
fermion ultracold gases the energy  of the diatomic molecule can be tuned (by means of a magnetic field)
above and below the continuum threshold has driven to the realization of the Bose-Einstein condensation
(BEC) in boson gases and of the BCS-BEC crossover in fermion gases driven by the Feshbach
resonance.\cite{48,49,50}  Therefore the \emph{shape resonance} and \emph{Feshbach resonance} belong to
the class of Majorana-Fano-Feshbach resonances due to configuration interaction effects between open
and closed scattering channels described by the Feshbach-Fano partitioning method.\cite{51,52,53}
In a multiband superconducting system, when the chemical potential is tuned at the Lifshitz transition
in one of the bands, the pairs of electrons at the Lifshitz transitions are in a quantum critical point
where the group velocity of electrons goes to zero and they form bosonic pairs that can be considered
the equivalent of the Majorana quasi-stationary states or the diatomic molecule in the magnetic field
in ultra cold gases.

In superlattices, the \emph{shape resonance}\cite{15,15a,15b,18,39} in the superconducting gaps appears
when the Fermi energy of electrons in a band is tuned around the Lifshitz transitions\cite{15} in
a different band. There are several types of Lifshitz transitions:

type I) when the Fermi energy crosses the band edge, at
$E_{edge}$, with the appearance or disappearance of a new Fermi surface (FS);

type II) when the Fermi energy crosses the electronic topological transition
${ETT}_{3D-2D}$ where one FS changes from 3D to 2D (i.e., from ``spherical'' to ``cylindrical'' ) or vice
versa, with the opening or closing of a neck in a tubular FS;

type III) at the singular point where the FS changes from 2D to one-dimensional (1D) or vice versa, with
the change of FS topology from a closed FS circle to disconnected Fermi arcs.\cite{35}
Once the heterostructure of materials selected as the building
blocks for the superconducting units and for the spacer layers has been synthesized
with its structural parameters, the tuning of the Fermi level at an ETT can be controlled by means of:
i) the ``charge transfer'' between the superconducting layers and the ``spacers''; ii) the
superlattice misfit strain\cite{56,57,58} between the superconducting and spacer layers; iii) the
thickness of the spacers; iv) the ordering of dopants in the spacer; v) the superstructures in
the superconducting layers forming stripes; vi) pressure.

\section{Breakdown of the standard BCS approximations}

The heterostructures at the atomic limit of superconducting units, where the chemical potential is tuned
at a type I Lifshitz transition (appearing or disappering of a new FS spot) and a the type
II Lifshitz transition (opening a neck in a FS), where the superconducting gaps show a
\emph{shape resonance}, have been called \emph{superstripes}.\cite{59,61,62} From the theoretical point
of view, the \emph{shape resonance} implies the breakdown of two main approximations of the standard
BCS theory for superconductivity: 1) the infinite distance of the Fermi energy from the band edges, which
is reasonable whenever the energy cutoff for the pairing interaction, $\omega_0$, is much smaller than
the Fermi energy $E_F$ (as measured from the nearest singular point in the electron spectrum): in this
case the electron density of states (DOS) can be taken as nearly constant within an energy window of
the order of $\omega_0$; 2) the single band approximation, which is altogether reasonable in the
dirty limit, when impurity scattering mixes the various electron components yielding an effective
single-band system.

The situation in which the first approximation breaks down (e.g., when the Fermi energy falls near an
electronic topological transition) has been called the van Hove scenario,\cite{63,64,65,66,67} or the
Pomeranchuk instability scenario,\cite{68,69,70} belonging to the class of quantum phase transition
scenarios investigated for pairing in single band superconductivity.\cite{71,72,73,74} While these
scenarios have been proposed for cuprates assuming a single-band approximation, recently there is
increasing evidence for a multiband superconductivity also in
cuprates\cite{a,1,2,3,4,5,6,7,8,75,76,77a,77b,77c,77d,78,78a,25}

As far as the second approximation is concerned, as we have recalled above, the very concept of
multiband superconductivity is meaningful only in the clean limit. Indeed, in nearly all metals and
alloys several bands cross the Fermi level, but impurity scattering leads to a mixing of the electron
states. Therefore, the single-band description, which is adopted, e.g., in the standard BCS theory, is a
reasonable approximation in the dirty limit. However, several experiments in doped diborides have shown
that the clean limit is robust in heterostructures at the atomic limit where both the different parity
of the bands and their different spatial location forbid the mixing of the different electron states at
the Fermi surfaces.\cite{15} Nowadays it is commonly accepted that the dirty-limit approximation
breaks down in magnesium diboride, in pnicitdes, in superlattices of carbon nanotubes,
in graphene and graphane layers, and multigap superconductivity may emerge also in
the presence of lattice disorder due to dopants and misfit strain of the
superlattice, when these are not apt to mix the different electronic components.

\begin{figure}[tpb]
\centering
\includegraphics [angle=0,scale=0.47]{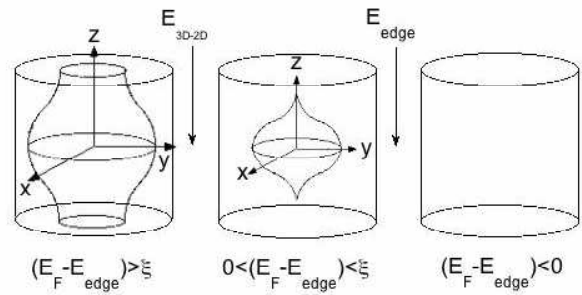}
\caption{Pictorial view of the evolution of the FS of the two-band
electronic system near a band edge crossing two Lifshitz transitions. The chemical potential is tuned so that $E_F$ crosses the band edge $E_{edge}$ of the second band (the type I Lifshitz transition) and the 3D-2D ETT at $E_{3D-2D}$ (the type II Lifshitz transition).  Here, $\xi$ is the transversal band dispersion.
On the left side a first large 2D FS coexists with a second tubular 2D FS; with decreasing
$E_F$ and crossing the energy $E_{3D-2D}$ where the second FS undergoes a 3D-2D ETT
changing its topology, the tubular 2D FS becomes a closed 3D FS as shown in the central panel;
only the first large 2D FS remains when the chemical potential is moved below the band edge
energy $E_{edge}$ of the second band as shown in the left panel.}\label{fig2}
\end{figure}

Another crucial feature of multiband superconductivity is that, beyond the standard
intraband attractive interactions that promote pairing within each band,
exchange-like interband interactions, that scatter Coopers pairs from one band to
another, become relevant. In the hypothetical absence of interband interactions, each
band would be characterized by its own superconducting critical temperature.
Arbitrarily weak interband interactions lead to a single critical temperature, like in the
proximity effect. It is important to notice that interband interactions are
generically repulsive and in this case interband pairing leads to condensate
wavefunction with opposite signs (the so called $s\pm$ pairing).\cite{15,16,22,22a,23}
In the \emph{shape resonance} scenario for multiband superconductivity, the
control of the ratio between the intensity of exchange-like interband pairing and
intraband Cooper pairing, by material design techniques, is crucial.

\section{Two band superconductivity at a band edge}

In this work we study the \emph{shape resonance} in the case when the
Fermi energy is tuned at a band edge energy $E_{edge}$ in one of the bands in a multiband superconductor.
In particular we focus on superlattices of metallic layers, with finite hopping probability between the
metallic layers. The universal feature of band edges in a superlattice of metallic layers is that the
small FS appearing or disappearing as the Fermi energy crosses the band edge has a 3D topology, due to
finite electron hopping between the layers separated by a finite potential barrier. We consider the
interesting case where the electron hopping between layers is so small that the transversal band dispersion
$\xi$ due to hopping between layers is much smaller than the energy separation between the subbands.  In
these superlattices the FS appearing or disappearing as the Fermi energy crosses the band edge has a
3D topology in the range $0<E_F-E_{edge}<\xi$. When the Fermi energy crosses the van Hove singularity,
an $ETT_{3D-2D}$ of the Lifshitz type called ``opening a neck'' takes place, with the FS changing
topology, from closed to tubular, at $E_F-E_{edge}=\xi$.

It has been shown\cite{15,17} that the optimum amplification of the critical
temperature occurs when $\xi$ is of the order of the energy cut-off of
the interaction, $\omega_0$. In this case, in the narrow energy range $3\omega_0$
where the chemical potential crosses both the band edge and the $ETT_{3D-2D}$, the condensate in the newly
appearing FS undergoes a crossover from a mixed Bose-Fermi regime to the Fermi BCS regime. For example,
in the regime
$-1<(E_F-E_{edge})/\omega_0<0$, the electron states associated with the newly appearing FS are unoccupied
in the normal state, and a BEC-like condensate of pairs with
bosonic character is formed below the critical temperature. In the range
$0<(E_F-E_{edge})/\omega_0<1$, all few electrons in the newly appearing 3D FS in
the normal state, condense below $T_c$, with the breakdown of the standard
approximation
$(E_F-E_{edge})/\omega_0\gg 1$. In the range $0<(E_F-E_{edge})/\omega_0<1$, in the
case under discussion $\xi=\omega_0$, the new FS has a 3D topology.
The ${{ETT}_{3D-2D}}$ plays a key role in the configuration interaction between
pairing channels in different bands in the range $0<(E_F-E_{edge})/\omega_0<2$.

We are interested in the evolution of superconducting gaps in the two bands and of the
critical temperature in the narrow energy range where the Feshbach-like \emph{shape
resonance}
in superconducting gaps takes place. So, we consider a first large 2D FS in the
standard BCS approximation (i.e., where the standard BCS approximation is
valid, large $E_F/\omega_0$ ratio, i.e., $E_F$ far from band edges) that coexists with
a second appearing small FS, where the standard approximation breaks down
(small $E_F/\omega_0$ ratio, with $E_F$ close to both the band edge and the
${{ETT}_{3D-2D}}$), see Fig. 2.

This simple model is apt to capture the physics emerging from experiments in doped
diborides,\cite{36} cuprates,\cite{78,78a} and pnictides.\cite{79,80,81,82,83,84,85}
Moreover it represents the optimal case to obtain the enhancement of critical
temperature via the \emph{shape resonance} mechanism.\cite{15} It
describes the multiband pairing at the bottom or top of 2D bands in a generic superlattice of layers. Here,
the effective potential barrier and the thickness of the spacer layers are selected to adjust the electron
hopping between superconducting layers and
the transversal energy dispersion $\xi$ of the order of the energy cut-off $\omega_0$ of the interaction.

The \emph{shape resonance} is determined by the relative strength of the
intraband coupling constants and the interband coupling constant determined by the type of material forming
the superconducting layers and the type of material forming the spacers. We consider here the case, typical
of pnictides, where the intraband attractive coupling strength is weak in both the first and second subband.
We have obtained the evolution of the superconducting gaps from below the edge, in the
mixed Fermi-Bose regime, to above the ETT in the Fermi-like regime. The direct evidence for the quantum
interference effect between pairing channels is provided by minima in the gap parameter for electrons in the
large FS. We report the evolution of this scenario with increasing exchange interband pairing. We also
investigate the effect of the variation of the transversal dispersion  $\xi$ that can be changed by changing
the separation space between layers. These results can explain the differences among 1111 pnictides,
with changing the rare earth ionic radius, and the difference between 1111 and 122 pnictides, where
the spacer layers changes in the superlattice, while the superconducting layers are kept fixed.

Finally, we show that the plots of the BCS gap ratios versus
the critical temperature that display significant changes in the different
superconducting regimes of the Feshbach-like \emph{shape resonance}. These plots can be easily compared
with experimental data and provide a very good experimental test for the theory of the Feshbach-like
\emph{shape resonance}.

There is a clear analogy between the Bose-Fermi crossover case studied here and the BEC-BCS crossover in
ultracold Fermi gases.\cite{perali04} The theory presented here follows Blatt's approach\cite{30}, and it
is appropriate for dealing with the case of a chemical potential near a band edge since the equation for
the gaps is solved together with the equation for the chemical potential in the superfluid phase. In fact
like for the BCS-BEC crossover, as shown by Leggett\cite{Leggett} the BCS wave function corresponding to
an ensemble of overlapping Cooper pairs at weak coupling considering a contact interaction (BCS regime)
evolves to non-overlapping pairs with bosonic character as the density decreases at the band edge (BEC
regime). The BCS theory remains valid in this limit if the BCS equation for the gap is coupled to the
equation that fixes the fermion density so that the chemical potential $\mu$ results strongly renormalized
below the critical temperature with respect to the Fermi energy $E_F$ of the non interacting system.

In ultracold gases the energy of the bound state of the diatomic molecule above or below the continuum is
tuned by a magnetic field. In the present case of superlattices the chemical potential can be tuned near
the ETT of a narrow band, for example by superlative misfit strain.

\begin{figure}[tpb]
\centering
\includegraphics [angle=0,scale=0.52]{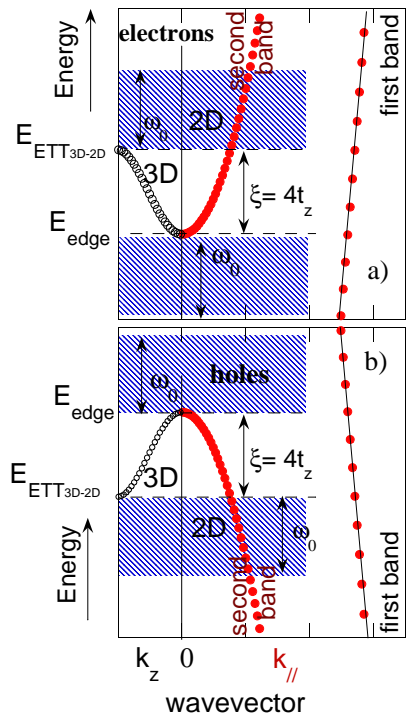}
\caption{(Color online). The shape resonance occurs by tuning the Fermi energy near
the band edge $E_{edge}$ of a second  electron-like band (upper panel) or a second hole-like band (
lower panel).
The second band coexists with a large first band (with the Fermi energy far from
its edges) that has a free-electron like dispersion. The second band has a free-electron dispersion in the
direction of the electron wavevector parallel to the plane ($k_{//}$) while in the perpendicular direction
($k_z$) of the superconducting layers it is determined by a periodic potential barrier determined by the
superlattice. This second band has a 3D character in the energy range between the band edge, and the van
Hove singularity energy $E_{{ETT}_{3D-2D}}$. The band dispersion is $\xi$=$4t_z$, where $t_z$ is the
electron hopping integral between the layers.}\label{fig3}
\end{figure}

\section{The two band model in a superlattice of metallic layers}

As pointed out above, the standard multiband BCS theory in the weak-coupling limit (or
the Eliashberg theory in the strong-coupling limit)\cite{22,22a,23} assumes a DOS
which is (at least approximately) constant within a window of thickness
$2\omega_0$ around the Fermi energy. This assumption is certainly not valid if
the chemical potential is located near a band edge or close to a van Hove singularity of
a superlattice of metallic layers.

Near the band edges in the $\ell$-th band, where $\nabla_{\mathbf k}E_\ell=0$, the
energy of an electron can be approximated by a free electron dispersion law
$E_{\ell,\mathbf k}=E_{\ell}+(\mathbf k^2/2m_\ell)$, where $m_\ell$ is the electron
effective mass at the band edge. Therefore, one can use the free electron
approximation
in a narrow region around the band edge, in the case when the band width is much larger
than $2\omega_0$, the energy range of interest for pairing. In the case of interest
here, however, while this approximation is valid in the $x,y$ plane of the
superconducting layers, it is certainly not valid in the $z$ direction, where
the dispersion is on the order of $\omega_0$. In this case, an anisotropic band
with weak dispersion in the $z$ direction and larger dispersion in the $x,y$ plane
above the band edge of the second band should be adopted,
\begin{equation}
E_{2,\mathbf k}^{3D}=E_2+E({k_z})+\frac{k_x^2+k_y^2}{2m_{//}}
\label{1}
\end{equation}
where $E(k_z)$ is the actual energy dispersion in the periodic potential of the
superlattice and $m_{//}$ is the effective mass in the $x,y$ plane. This situation
is obtained within a model where a free electron gas is confined in a potential which
is periodic in the $z$ direction,
\begin{equation}
\mathcal{W}(z)=\sum_{n=-\infty}^{n=+\infty}\mathcal{W}_b(z-nd),
\label{2}
\end{equation}
where $\mathcal{W}_b(z)=-V_b$ for $|z|\le L/2$ and $\mathcal{W}_b(z)=0$ for
$L/2<|z|<d/2$, $L$ is the width of the confining well and $d$ is the periodicity of
the superlattice in the $z$ direction. This periodic potential mimics the
phenomenology, e.g., of the pnictides, diborides, and stacks of graphene layers made
of a superlattice of stacked planes, as shown in Fig. 1. The confining potential generates
a band structure organized in subbands. Each subband has a dispersion in the
transversal direction, as shown in Fig. 2, that gives a Fermi surface with 3D
topology, with closed isoenergetic surface, near the lower band edge $E_{edge}=E_{\ell,L}$,
and a 2D character, with isoenergetic surfaces open in the $z$ direction, above some
energy threshold $E_{ETT_{3D-2D}}=E_{\ell,T}$. The present model is apt to describe
quantum interference phenomena between different scattering channels in a large and a
small FS which are the object of this work, when the chemical potential $\mu$ of the
system is tuned near the bottom of a $\ell$-th subband, within a window of width
$4\omega_0$. In the energy window of width $2\omega_0$ the first band, with a large 2D
tubular FS, has a constant DOS $N_1$. The second
FS appears as the Fermi energy crosses the level $E_{edge}=E_{\ell,L}$, and changes
from closed 3D to tubular 2D topology as the Fermi energy crosses the level
$E_{ETT_{3D-2D}}=E_{\ell,T}$ as it is shown in Fig. 2. The model for electron-like
FS can be easily extended to hole-like FS as shown in Fig. 3. The DOS of the second
FS $N_2$ near the edge has the typical 3D behavior.

In order to solve the BCS equations in the shape resonance scenario it is necessary to
determine the electron wavefunction. Using the model of a free electron gas confined
in a superlattice, the wavefunction of electronic states can be obtained by solving
the Schr\"{o}edinger equation, so that we can calculate the anisotropic $\mathbf{k}$-
dependent gaps in the wavevector space and the interference between the intra-band pairing
channels and the pair exchange determined by the interband pairing terms.
The solution of the Schr\"{o}edinger equation for the 1D periodic potential of the
superlattice of layers allows us to calculate the gaps also in correspondence of the
3D to 2D electronic transition $ETT_{3D-2D}$ at the van Hove singularity, beyond the
standard BCS approximations.\cite{30,35}

Going from a single slab\cite{30} to a 3D superlattice\cite{35}, a 3D condensate is
formed, reducing the effect of fluctuations of the superconducting order parameter,
which suppress the mean-field superconducting $T_c$. The crossover from 2D to 3D can be
described in our model by changing from an infinite potential barrier between the
planes of the superlattice in the $z$ direction, provided by spacer layers, to a finite
barrier ($V_b$). This yields a finite hopping term that broadens the sharp discontinuity of
the DOS of a pure 2D band. This broadening increases the width of the shape resonance and,
at the same time, yields a 3D condensate. Moreover, it is possible to design
artificial superlattice heterostructures at optimum shape resonance condition, i.e.,
where the value of the potential barrier $V_b$ and its width are such that the subband
dispersion in $z$ direction $\xi$ is of the order of energy cut-off $\omega_0$ of the
interaction. Therefore, it is possible to tune the system to the optimum shape
resonance condition, achieving an enhancement of the critical temperature.

The intrinsic $\mathbf{k}$-dependence of the pairing interaction
$\mathcal{V}_{\mathbf{k},\mathbf{k}'}^{\ell,\ell'}$ in the superlattice with
wave vector $\mathbf{k}_z$ induces a structure in the $\mathbf{k}$-dependent
interband coupling interaction for the electrons that determines the quantum
interference between electron pairs wavefunction in different subbands of the
superlattice.\cite{15} We calculate the term
$\mathcal{V}_{\mathbf{k},\mathbf{k}'}^{\ell,\ell'}$ following our previous work for a
superlattice of wires,\cite{35} determining the matrix elements of local interaction
potential and then introducing a cutoff with two $\theta$ functions,
\begin{equation}
\mathcal{V}_{\mathbf{k},\mathbf{k}'}^{\ell,\ell'}=
\widetilde{
\mathcal{V}}_{\mathbf{k},\mathbf{k}'}^{\ell,\ell'}
\theta(\omega_0-|\xi_{\ell,\mathbf{k}}|)\theta(\omega_0-|\xi_{\ell',\mathbf{k}'}|)
\label{3}
\end{equation}
where $\mathbf{k}=\mathbf{k}_z$ ($\mathbf{k}'=\mathbf{k}'_z$) is the superlattice
wavevector, in the $z$ direction, perpendicular to the planes, of the initial (final)
state in the pairing process, and
\begin{equation}
\widetilde{
\mathcal{V}}_{\mathbf{k},\mathbf{k}'}^{\ell,\ell'}=
\frac{c_{\ell,\ell'}}{N_0(E_F)V_{3D}}I_{\mathbf{k},\mathbf{k}'}^{\ell,\ell'},
\label{4}
\end{equation}
where $N_0(E_F)$ is the DOS at $E_F$ for a free electron 3D system, $V_{3D}$ is the
volume of the system,
\begin{equation}
I_{\mathbf{k},\mathbf{k}'}^{\ell,\ell'}=-d\int_{d}
\psi_{\ell,-\mathbf{k}}(z)\psi_{\ell',-\mathbf{k}'}(z)
\psi_{\ell,\mathbf{k}}(z)\psi_{\ell',\mathbf{k}'}(z)dz
\label{5}
\end{equation}
and the $\psi_{\ell,\mathbf{k}}(z)$ are the eigenfunctions in the superlattice of
quantum wells, normalized so that
$\int_{d}dz|\psi_{\ell,\mathbf{k}}(z)|^2=1$.
The use of single cut off in two band superconductivity has been justified in detail
in ref. \cite{Entel}

The dimensionless factor $c_{\ell,\ell'}=(-1)^{\delta_{\ell,\ell'}}c^0_{\ell,\ell'}$
assumes positive values for $\ell=\ell'$ (intraband Cooper pairing) and negative
values for $\ell\neq\ell'$ (repulsive exchange-like interband pairing, with
$c_{\ell,\ell'}=c_{\ell',\ell}$) and measures the relative intensity of intraband
and interband pairing strength. In fact, it multiplicates the $\mathbf{k}$-dependent
integral and therefore permits to simulate the behavior of different
superconductive multilayer compounds controlling the ratio between intensities of
intraband and interband pairings.

\begin{figure}[tbp]
\centering
\includegraphics [angle=0,scale=0.49]{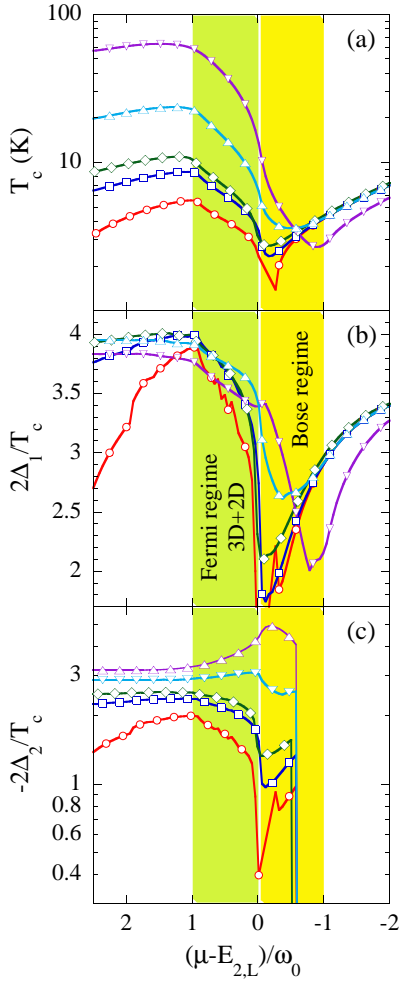}
\caption{(Color online). The case of weak coupling in the second band (the so called
pnictide case) where the intraband coupling parameters ratio is fixed at
$c_{2,2}/c_{1,1}=0.45$. The critical temperature $T_c$  (panel a), the ratio
$2\Delta_1/T_c$ (panel b) and the ratio $-2\Delta_2/T_c$ (panel c) are plotted as function of the
reduced Lifshitz parameter $(\mu-E_{2,L})/\omega_0$. The different curves in each panel represent the
cases of different interband pairing strength: the case for $c_{12}/c_{11}=-2.73$ (violet curve with
open triangles); the case for $c_{12}/c_{11}=-1.59$ (light blue curve with open triangles);  the case for
$c_{12}/c_{11}=-1.04$ (green curve with open diamonds); the case for $c_{12}/c_{11}=-0.91$ (dark blue curve
with open squares) and shows the case for $c_{12}/c_{11}=-0.68$ (red curve with open circles).
For the largest case of interband pairing strength $c_{1,2}$ considered here,
the critical temperature reaches 300K in the range $1<(\mu-E_{2,L})/\omega_0<2$.
The shape resonance antiresonance appears as a minimum in the critical temperature for $(\mu-E_{2,L})/
\omega_0$=-1 for higher interband repulsive interaction and moves towards
$(\mu-E_{2,L})/\omega_0=0$ for the smaller interband repulsive interaction.
The ratio $2\Delta_1/T_c$ for the first band shows maxima and minima like the critical temperature. On
the contrary the ratio $2\Delta_2/T_c$ shows a minimum at $(\mu-E_{2,L})/\omega_0=0$ and it has a
maximum at the value of maximum $T_c$ for weak interband repulsive coupling, on the contrary it exhibits
a maximum in the Bose-like regime(yellow region) $(0>\mu-E_{2,L})/\omega_0>-1$ and a  minimum value in correspondence
of the maximum $T_c$ for very strong interband repulsive pairing.}\label{fig4}
\end{figure}

In order to determine the gaps and chemical
potential self-consistenly and to calculate the superconductive $T_c$ we use iterative
solving methods for the coupled BCS-like equations
\begin{equation}
\Delta_{\ell,\mathbf{k}}(\mu)=-\frac{1}{M}\sum_{\ell',\mathbf{k}'}
\frac{\mathcal{V}_{\mathbf{k},\mathbf{k}'}^{\ell,\ell'}\Delta_{\ell',\mathbf{k}'}}{2
\sqrt{(E_{\ell',\mathbf{k}'}-\mu)^2+\Delta_{\ell',\mathbf{k}'}^2}},
\label{6}
\end{equation}
\begin{equation}
\rho=\frac{1}{d^2}\sum_{\ell,\mathbf{k}}\left[1-
\frac{E_{\ell,\mathbf{k}}-\mu}{\sqrt{(E_{\ell,\mathbf{k}}-\mu)^2+
\Delta_{\ell,\mathbf{k}}^2}}\right],
\label{7}
\end{equation}
starting with an initial gap parameter equal to a constant and an initial chemical
potential equal to the Fermi level in the normal state and considering the convergence occurred for
relative variation of the gap and charge density $\rho$ less than $10^{-6}$. Here $M$ is the total
number of wavevectors $\mathbf{k}'$ and $\rho$ is the electron density. The superconducting
critical temperature $T_c$ is calculated by iterativly solving the linearized equation
\begin{equation}
\Delta_{\ell,\mathbf{k}}=-\frac{1}{2M}\sum_{\ell',\mathbf{k}'}
\mathcal{V}_{\mathbf{k},\mathbf{k}'}^{\ell,\ell'}
\ \frac{\tanh\left(\frac{E_{\ell',\mathbf{k}'}-\mu}{2T_c}\right)}
{E_{\ell',\mathbf{k}'}-\mu}\Delta_{\ell',\mathbf{k}'},
\label{6}
\end{equation}
until the vanishing solution is reached with increasing temperature.
Here, the gaps depend on the superlattice wavevector $\mathbf{k}$ as well as on the
subband index. Hence, $T_c$ and the gap at a given point of $\mathbf{k}$-space become
implicit function of all the different values of the gaps in the entire
$\mathbf{k}$-space.

In the standard BCS theory, where the Fermi energy is far from the band edges,
the relative variation of the chemical potential going from the normal to the
superconducting state is expected to be negligible. This is not true when the
chemical potential is tuned near the band edge of the second band. In fact,
our calculation yields a significant variation of the chemical potential in
the superconducting phase, as a function of the charge density. A
relative variation of the chemical potential going from the normal to the
superconducting phase, as large as $10^{-3}$, is obtained near the band edge and at
the $ETT_{3D-2D}$, within a range $4\omega_0$. The variation starts to be large,
as compared with the standard BCS result, in proximity of the Bose-Fermi crossover
regime below the band edge up to well beyond the $ETT_{3D-2D}$.

The Feshbach-like shape resonance regime occurs properly in correspondence of
this large
variation of the chemical potential between the normal and superconducting phases.
Our theoretical approach provides a direct measure of the gap, at a given point of
$\mathbf{k}$-space anisotropy for both intraband and exchange interband terms. The
intraband distributions of the two bands show different shapes and widths and have
different range of values. The resulting matrix of coupling terms is obtained
exclusively from the eigenfunctions of the superlattice and is asymmetric.

Below, we present numerical results for the solution of the self-consistency
equations which determines the values of the gap in the large FS and in the small FS.
We discuss the behavior of $\Delta_1$ and $\Delta_2$, defined as average values of
the gaps on the corresponding branches of the Fermi surface, and of the critical
temperature $T_c$.

\begin{figure}[tbp]
\centering
\includegraphics [angle=0,scale=0.49]{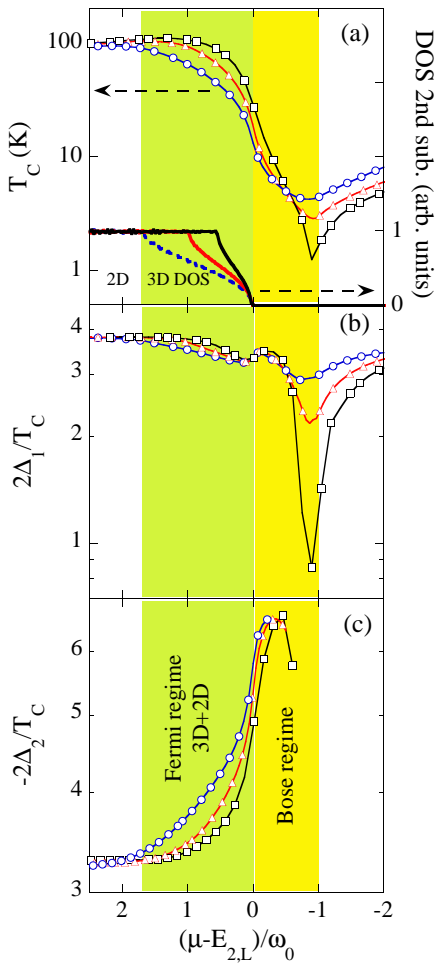}
\caption{(Color online). The effect of the variation of transversal dispersion on
the shape resonance, the case of weak coupling in the second band (the so called
iron pnictide case) for the case of strong interband repulsive interaction. The intraband and
interband coupling parameter ratio
($c_{2,2}/c_{1,1}=0.45$ and $c_{1,2}/c_{1,1}=$ -3.6, with $c_{1,1}=0.22$) are fixed.
The critical temperature (panel a), the ratios $2\Delta_1/T_c$ (panel b)
and $-2\Delta_2/T_c$ (panel c) as function of the reduced Lifshitz parameter
$(\mu-E_{2,L})/\omega_0$ are plotted for $\xi=110$ (blue curve with open circles), 68 (red curve
with open triangles) and 36 meV (black curve with open squares). This last case is close to the case of a
single quantum well. We also
plot the DOS of the second subband in the inset of panel (a). The ratio
$2\Delta_1/T_c$, probing the pairing channel in the first large FS, exhibits two
minima due to the negative interference effect (or antiresonance) typical of
shape resonances, at the band edge $(\mu-E_{2,L})/\omega_0$=0 and the the ETT
$(\mu-E_{2,L})/\omega_0$=-1. The antiresonance effect in the large FS increases
for decreasing dispersion in the second band. The value of the second superconducting
gap is nonzero even before the lower band edge $E_{2,L}$ is reached. In the second
band the maximum value of the ratio is always reached in the Bose regime and remains
the same, $-2\Delta_2/T_c\simeq6.7$, for all dispersion values.}\label{fig5}
\end{figure}

We discuss the case where the coupling term in the second band is smaller than in the first band
(called here the pnictide-like case). This case allows to emphasize the role of the exchange-like
interband pairing and to use the BCS weak limit for interactions for contact interactions,
neglecting retardation effects. In the extreme case when the intraband coupling in the second band is
zero, the pairs in the second band are formed only thanks to interband exchange-like pairing.

Therefore in this approach we consider the intraband coupling $c_{1,1}$ for the band that does not show
the Lifshitz transition, the intraband coupling $c_{2,2}$ for the one that shows the Lifshitz transition
and the inteband coupling $c_{1,2}$ for the exchange-like pairing.

Fig. 4 shows the case of weak coupling in the second band where the intraband coupling parameters ratio
is fixed at $c_{2,2}/c_{1,1}=0.45$. The critical temperature $T_c$  (panels a ),
the ratio $2\Delta_1/T_c$ (panel b), and the ratio $-2\Delta_2/T_c$ (panel c) are
plotted as function of the reduced Lifshitz parameter $(\mu-E_{2,L})/\omega_0$.
The different curves in each panel represent the cases of different interband
pairing strengths ($c_{1,2}/c_{1,1}=$-0.68, -0.91, -1.04, -1.59, -2.73). In correspondence of the largest
interband pairing strength examined here, $c_{1,2}=$-2.73, we have obtained critical temperatures as large
as 50K, in the range $1<(\mu-E_{2,L})/\omega_0<2$.
The antiresonance in the first gap appears when the ratio
$2\Delta_1/T_c$ reaches the minimum value, that can be much smaller than standard BCS
value ($2\Delta/T_c=3.53$). The antiresonance appears in the range $-1<(\mu-E_{2,L})/\omega_0<0 $ and moves
from -1 to zero decreasing the interband pairing.
Moreover the second superconducting gap values results to be nonzero even before the
lower band edge $E_{2,L}$ is reached. We notice that the maximum $T_c$ is reached in the BCS regime zone
near the type II Lifshitz transition ($ETT_{3D-2D}$).

Iron pnictides superconductors are certainly more three dimensional than cuprates
(they have a quasi-2D electronic structure), but the dimensionality of electronic structure in these compounds
is also less clear, in fact
even if some band structure works predict strong k$_z$ dispersion,\cite{101}
other works predict weak dispersions in AFe$_2$As$_2$ and ReOFeAs,\cite{102}
determined mainly by Fe d-orbitals of FeAs layers.
Furthermore many angle-resolved photoemission spectroscopy experiments\cite{103,104,105,106,107,108}
reveal weak k$_z$ dependences for the bands electronic structure,
in fact most of the band dispersion of 1111
and 11 iron pnictides, like LaOFeAs and NaFeAs,\cite{109,110}
are in the range of $20\sim40$ meV.
In 122 undoped pnictides, like BaFe$_2$As$_2$, the k$_z$ dispersion is
rather small,\cite{111} although in BaFe$_{2-x}$Co$_x$As$_2$, with increasing Co concentration,
the k$_z$ dispersion increases up to 120 meV.\cite{112}

Fig. 5 shows the effect of the variation of out of plane dispersion on the
shape resonance, in the case of weak coupling in the second band (the so called iron
pnictide case) for strong interband pairing.
The DOS of the second subband for different dispersions
($\xi =110$, 68 and 36 meV), as a function of the reduced Lifshitz parameter
$(\mu-E_{2,L})/\omega_0$, is shown as an inset in panel (a). The intraband and
interband coupling parameter ratio ($c_{2,2}/c_{1,1}=0.45$ and $c_{1,2}/c_{1,1}=$
-3.6, with $c_{1,1}=0.22$) are fixed. The critical temperature, panel (b), the ratios
$2\Delta_1/T_c$, panel (c), and $-2\Delta_2/T_c$, panel (d), are plotted as a function
of the reduced Lifshitz parameter $(\mu-E_{2,L})/\omega_0$. The minima of the ratio
$2\Delta_1/T_c$, probing the pairing channel in the first large Fermi surface show
two minima due to the negative interference effect or antiresonance typical of
shape resonances at the band edge $(\mu-E_{2,L})/\omega_0$=0, and the ETT
$(\mu-E_{2,L})/\omega_0$=-1. The antiresonance effect in the large FS increases for
decreasing dispersion in the second band. The value of the second superconducting gap
is nonzero even before the lower band edge $E_{2,L}$ is reached. In the second band,
the maximum value of the ratio is always reached in the Bose regime and remains the
same, $-2\Delta_2/T_c\simeq6.7$, for all dispersion values.

\begin{figure}[tbp]
\centering
\includegraphics [angle=0,scale=0.49]{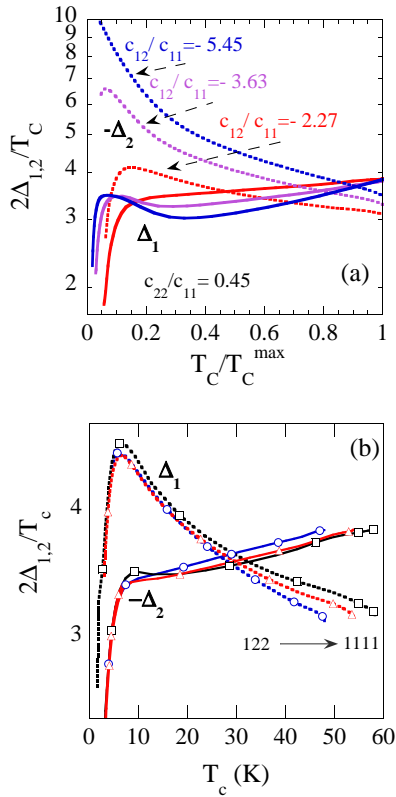}
\caption{(Color online).  (a) The ratio $2\Delta_1/T_c$ and $-2\Delta_2/T_c$ as
function of the ratio $T_c/T_c^{max}$, in the case of weak coupling in the second
band, with fixed $c_{2,2}/c_{1,1}=0.45$, $c_{1,1} = 0.22$ and interlayer dispersion
($\xi/\omega_0= 1$), for three cases of different interband coupling
parameters ($c_{1,2}/c_{1,1}=$ -2.27, -3-63, and -5.45) ratio. (b) The
ratios $2\Delta_1/T_c$ and $-2\Delta_2/T_c$ as functions of $T_c$ in the case of
weak coupling in the second band with fixed $c_{2,2}/c_{1,1}=0.45$, $c_{1,1} = 0.22$
and interlayer dispersion ($\xi/\omega_0= 1$), and in the case of
weak coupling in the second band with fixed $c_{1,2}/c_{1,1}=0.45$, $c_{1,1} = 0.22$
and interlayer dispersion $\xi= 110$ (blue curve with open circles), 68 (red curve with
open triangles) and 36 meV (black curve with open squares).}\label{fig6}
\end{figure}

Using STM spectroscopy it is now possible to measure the gaps and the critical temperature in the same
set of experiments. Therefore the BCS gap ratios $2\Delta_1/T_c$ and $-2\Delta_2/T_c$ as functions
of the ratio $T_c/T_c^{max}$ can be measured directly for different samples of different gate voltages
tuning the chemical potential. The present theory is able to predict these curves and we show that these
curves provide a direct measure of the relevance of interband coupling versus Cooper pairing. In fig. 6
we show the case strong interband repulsive interaction and weak coupling for Cooper pairing.
We keep constant the ratio $c_{2,2}/c_{1,1}=0.45$, $c_{1,1} = 0.22$ and $\xi/\omega_0= 1$ and we present
the expected behavior of these curves for strong interband interaction. The BCS Gap ratio parameters
are plotted for three cases of different interband coupling parameters. When the interband repulsive
term is dominant, the BCS gap ratio is the same in both bands at the maximum critical temperature.
The difference between the gap ratios diverges decreasing the critical temperature. This behavior is
very similar to case of cuprates in fact in the underdoped region where the critical
temperature goes to zero and while a first BCS-like gap
decreases with the decreasing critical temperature, the second gap (called
pseudogap) increases by decreasing the critical temperature.

We have investigated the effect of the band dispersion for a case on moderate relevance of the
interband  pairing such that there is a characteristic critical temperature where the two gaps are the
same and near the standard BCS gap parameter 3.53. We can see that going from  the case where the
band dispersion is low, i.e., we are close to the regime for a single isolated slab of Blatt to the
regime where the dispersion is as large as twice the energy cutoff of the interaction there
minor variations. This result shows that the shape resonance due to quantum confinement is a robust
feature also for appreciable electron hopping between the quantum wells.

\section{Conclusions}

The present work provides an interpretation for the properties of quantum size
effects in superconducting multilayers and indicate a possible roadmap for the discovery of novel HTS
like graphene bilayers and graphane superlattices that should share similarities with the known multigap
HTS families where the \emph{shape resonance} is driven by the interband pairing mechanisms.

Here, we have investigated the effect of electron hopping between quantum wells that can be changed by
changing the thickness of spacer layers, like in 1111 pnictides  where the rare earth ionic radius in
the spacer layer can be changed, or by changing the type of material  forming the spacers like going
from 1111 to 122 pnictides. To take into account this effect on the shape resonance, we changed the
dispersion $\xi$, while keeping it of the order of the pairing interaction energy cut-off $\omega_0$.
We have fixed in our calculation the intraband Cooper pairing much
weaker than the exchange-like interband repulsive interaction. This is clearly the case of iron pnictides
that exhibit critical temperatures of the  order of 50K mostly driven by interband pairing.
Namely, we have examined the case of intraband and
interband coupling parameter ratio $c_{2,2}/c_{1,1}=0.45$ with variable interband
repulsive interactions and variable single electron dispersion in the transversal
direction.

The ratios $2\Delta_1/T_c$ and $-2\Delta_2/T_c$ predicted for high-$T_c$
iron-based layered compounds and cuprates as functions of $T_c$  show that for the
highest $T_c$ samples the gaps become equivalent while they are quite different in the low $T_c$ samples,
that appears to be the scenario for cuprates and for pnictides.
The antiresonance typical of shape resonance occurs where the Lifshitz energy parameter is zero (i.e appearing of a new Fermi surface spot) for weak interband exchange pairing and shift to -1 for a very strong interband exchange pairing. The relevant result is that the maximum of the critical temperature appears near the  Lifshitz transition of the type opening a neck in corrugated cylindrical surface therefore it moves with the variation of the electronic hopping between the layers i.e., the energy dispersion in the normal direction respect to the superconducting layers.

\begin{acknowledgments}
We thank Andrea Perali, Ilya Eremin, and Andrei Shanenko for useful discussions.
\end{acknowledgments}

\end{document}